\documentclass[sigconf, nonacm]{acmart}
\usepackage{subcaption}
\usepackage{pifont}

\usepackage{fancyvrb}
\usepackage{algorithm}
\usepackage{algpseudocode}
\usepackage{enumitem}
\usepackage{amsmath}
\usepackage{graphicx}
\usepackage{tikz}
\usetikzlibrary{positioning}

\usepackage{tabularx}
\usepackage{float}
\usepackage{caption}
\usepackage{bbding}

\newcommand{\squishlist}{%
  \begin{itemize}[leftmargin=1.5em, itemsep=2pt, parsep=0pt, topsep=3pt, partopsep=0pt]
}
\newcommand{\squishend}{\end{itemize}}
\newcommand{\stitle}[1]{\noindent\textbf{#1}\ }
\newcommand{\lionrockmark}{\ensuremath{{}^{\dagger}}}
\newcommand{\whumark}{\ensuremath{{}^{\ddagger}}}
\newcommand{\cuhkszmark}{\ensuremath{{}^{\S}}}
\newcommand{\corrmark}{\ensuremath{{}^{*}}}
\newcommand{\authsep}{,\ }
\begin{document}

\title{BatchWeave: A Consistent Object-Store-Native Data Plane for Large Foundation Model Training}

\author{Ting Sun\lionrockmark\authsep Junjie Zhang\lionrockmark\authsep Xiao Yan\whumark\authsep Songxin Zhang\lionrockmark\authsep Zhuoyang Song\lionrockmark\\
Jingyi Xi\lionrockmark\authsep Zunyao Mao\lionrockmark\authsep Bingyi Jing\cuhkszmark\authsep Jiaxing Zhang\lionrockmark\corrmark\authsep Zejian Xie\lionrockmark\corrmark}
\affiliation{%
  \institution{\lionrockmark Lionrock AI Lab, China Merchants Group, Hong Kong, China\\
  \whumark Wuhan University, Wuhan, China; \cuhkszmark The Chinese University of Hong Kong, Shenzhen, China\\
  \texttt{\{sunting2,junjayzhang,zhangsongxin,songzhuoyang\}@cmhk.com}\\
  \texttt{\{xijingyi,maozunyao,zhangjiaxing,xiezejian\}@cmhk.com}\\
  \texttt{yanxiaosunny@whu.edu.cn; bingyijing@cuhk.edu.cn}\\
  \corrmark Corresponding authors}
  \city{}
  \country{}
}
\renewcommand{\shortauthors}{Sun et al.}

\begin{abstract}
Modern Large Foundation Model (LFM) training has transformed the data pipeline
from a static ingestion layer into a dynamic component that must co-evolve with
the training process. Existing systems are ill-equipped: colocated dataloaders
offer no failure isolation, while message queue-based disaggregated dataloaders
operate on a record/offset abstraction that cannot express the batch-level
semantics required by distributed training.
 
We present BatchWeave, an object-store-native training data plane for distributed
LFM training. BatchWeave uses versioned manifests and conditional object writes
to coordinate batch publication, recovery, and lifecycle management. First, it
introduces the Transactional Global Batch (TGB), which builds on
versioned-manifest ACID storage semantics and extends them with
training-specific consistency, including atomic all-rank batch visibility, a
globally ordered step sequence, checkpoint-aligned lifecycle management, and
end-to-end exactly-once recovery. Second, it realizes recovery and retention
directly in the storage layer, by durably persisting producer state through
the commit protocol and tying reclamation to distributed checkpoint state. Third, its Decentralized
Adaptive Commit (DAC) algorithm sustains stable ingestion throughput as the
manifest grows, without any inter-producer communication.
 
Evaluations on large-scale multimodal pre-training and SFT workloads using 64
GPUs show that BatchWeave outperforms colocated dataloader throughput while providing
full failure isolation, outperforms Apache Kafka in ingestion throughput, and achieves
lower consumer read latency than Kafka.
\end{abstract}

\maketitle

\section{Introduction} \label{sec:intro}

Data loading has long been a well-studied component of model training. In
conventional deep learning workloads, the data pipeline is straightforward: a
static, pre-processed dataset is shuffled offline, partitioned into fixed-size
batches, and streamed to the training process at a predictable rate. Systems such
as tf.data~\cite{murray2021tfdata} and MosaicML Streaming~\cite{mosaicmlstreaming}
are designed around this model, optimizing for throughput and shuffle quality over
a fixed corpus. To prevent preprocessing from bottlenecking GPU utilization,
disaggregated variants offload preprocessing to independent worker pools: tf.data
service~\cite{audibert2023tfdata_service} and Cachew~\cite{graur2022cachew} run
preprocessing on separate CPU nodes and deliver batches to training ranks via RPC,
while CoorDL~\cite{mohan2021coordl} coordinates cache sharing across colocated
training jobs. These systems improve resource efficiency, but all share a common
assumption: the dataset is static, batch membership is determined before training
begins, and batches are ephemeral delivery artifacts that are discarded once
consumed.

Large Foundation Model (LFM) training breaks these assumptions. The data pipeline
is no longer a passive feeder of a fixed corpus but a dynamic, runtime-coupled
component that must co-evolve with the training process. It is dynamic in three
aspects that existing systems are not designed to handle.

\squishlist

\item \textbf{Data volume.} In multimodal pre-training and supervised fine-tuning
with video or interleaved image-text data, raw inputs undergo heavy runtime
preprocessing before reaching the optimizer. Video decoding, frame extraction,
resolution rescaling, and token packing all introduce data expansion whose volume
depends on the input content and current model configuration. Intermediate data can
exceed the raw dataset size by one to three orders of magnitude. This unpredictable
expansion makes static resource allocation infeasible and demands that preprocessing
be decoupled from training, with elastic storage to absorb runtime-materialized
data.

\item \textbf{Batch membership.} In workloads that use online token packing or
model-dependent sample selection (such as long-context continued pre-training or
reinforcement learning from human feedback), the preprocessing pipeline itself runs
at training time, and the exact composition of each batch is determined by its
output. Batch boundaries are known only after preprocessing completes; no static
partitioning scheme can capture them in advance. The data plane must therefore treat
batch membership as a runtime artifact and provide a mechanism to expose complete
batches atomically to all training ranks.

\item \textbf{Checkpoint alignment.} Modern LFM pipelines interleave pre-training,
fine-tuning, and reinforcement learning stages, each of which may require rolling
back to an earlier checkpoint or replaying a specific batch sequence. Rather than a
form of dynamism per se, this reflects a requirement for persistence and
versioning: the data plane must maintain a durable, checkpoint-aligned batch history
so that any live checkpoint can deterministically recover the exact data it consumed.
Ephemeral delivery systems cannot provide this guarantee.

\squishend

These requirements call for a training-aware data plane that understands the
semantic structure of training steps, not one that merely transports individual
records. Existing solutions do not meet this bar. Colocated dataloaders provide
natural semantic alignment but no failure isolation: a preprocessing crash stalls
the entire job. They are also subject to structural resource contention, where
preprocessing threads share CPU cycles and memory bandwidth with the training
process on the same node, imposing a throughput ceiling that in-rank engineering
cannot overcome under heavy preprocessing workloads. Disaggregated dataloaders
with centralized services (tf.data service, Cachew) decouple preprocessing from
training and support elasticity, but deliver batches ephemerally via RPC with no
persistent history and no batch-level atomicity guarantees. Neither category can
express the training-aware semantics that LFM workloads require: a batch spanning
multiple dynamically produced objects must be either fully visible or fully
invisible to all data-parallel ranks at once, and its membership must be
recoverable across failures and rollbacks.

Object storage is a natural substrate for a disaggregated training data plane. Its
append-only write model, elastic scalability, and decentralized access require no
dedicated ingestion-service provisioning. Inspired by versioned table
architectures~\cite{delta,iceberg,lance}, we propose to govern data visibility
through versioned metadata manifests rather than record offsets, separating the
act of writing data from the act of exposing it to consumers. A new training job
requires only a fresh namespace prefix, with no partition assignment and no
cold-start overhead. Object storage alone, however, provides only single-object
atomicity: it offers no batch-level transactional semantics and no
training-progress-aware lifecycle management. More fundamentally, a central
coordinator is structurally misaligned with LFM training: preprocessing pools are
elastic, training jobs require flexible scaling of producers and consumers
independently, and any server-side component introduces a new failure domain and
provisioning overhead orthogonal to the job. BatchWeave's design principle is
therefore that object storage and versioned manifests should be the shared
substrate for publication, recovery, and lifecycle management.

Doing so requires solving problems at three levels.

\squishlist

\item \textbf{Concurrent commit throughput.} Decentralized writes rely on
Optimistic Concurrency Control (OCC) over a versioned manifest. Under high producer
concurrency and a monotonically growing manifest, the system must balance commit
conflicts, commit overhead, and ingestion throughput with no inter-producer
communication.

\item \textbf{Batch-level semantics.} The system must guarantee atomic visibility
of each training batch across all data-parallel ranks, a globally consistent batch
ordering, and exactly-once semantics under producer failures and consumer restarts.

\item \textbf{Checkpoint-coupled lifecycle.} Checkpoint-based rollbacks and
cross-run reuse require a versioned batch history, while training-progress-aligned
reclamation is needed to bound storage costs without a persistent coordination
service.

\squishend

We present \textbf{BatchWeave}, an object-store-native training data plane. Its
core abstraction is the \textbf{Transactional Global Batch (TGB)},
which treats each training batch as a first-class persistent entity with
atomicity, durability, and globally consistent ordering guarantees. BatchWeave materializes TGBs as versioned, manifest-referenced
structures on object storage, decoupling data production from training consumption
while preserving the semantic boundaries required by distributed optimization. It
pairs the TGB abstraction with checkpoint-aligned lifecycle management,
end-to-end exactly-once semantics, and the \textbf{Decentralized Adaptive Commit
(DAC)} algorithm, which regulates each producer's commit cadence based on online
estimates of commit contention and manifest growth, sustaining stable ingestion
throughput without any inter-producer communication.

The technical contributions of this paper are as follows.

\squishlist

\item \textbf{Object-store-native architecture.} We establish object storage and
versioned manifests as the substrate for batch visibility, fault recovery, and
lifecycle management. A new training run requires only a fresh namespace prefix,
and the critical path consists of object writes, conditional manifest
publication, and direct range reads (\S\ref{sec:overview}).

\item \textbf{Transactional Global Batch (TGB) abstraction.} We introduce the TGB
as a first-class persistent entity with atomicity, durability, and globally
consistent ordering, unifying data visibility with training step semantics. All data-parallel ranks observe a
globally consistent and atomically visible batch at every optimization step;
exactly-once delivery and checkpoint-aligned recovery are achieved entirely through
object store primitives (\S\ref{sec:design}, \S\ref{sec:control:fault}).

\item \textbf{Checkpoint-aligned lifecycle and DAC.} We design a lifecycle
management mechanism that ties data retention to distributed checkpoint progress,
supporting safe rollback, cross-run reuse, and bounded storage overhead. The
Decentralized Adaptive Commit (DAC) algorithm sustains stable ingestion throughput
as the manifest grows, without any inter-producer communication
(\S\ref{sec:control}).

\squishend

We evaluate BatchWeave on large-scale multimodal pre-training and SFT workloads
using 64 GPUs. BatchWeave outperforms the expert-optimized colocated pipeline by
2.68--7.73$\times$ in end-to-end throughput while providing full failure isolation,
outperforms Kafka in producer ingestion throughput as the number of producers
scales, and achieves lower consumer read latency than Kafka at all measured scales. DAC maintains stable throughput across long training runs where
fixed-interval strategies degrade due to manifest growth.

\section{Background} \label{sec:background}

\subsection{Requirements for LFM Training}
\label{sec:background:requirements}

\stitle{SPMD execution and the Global Batch.}
Modern LFM training follows the Single Program Multiple Data (SPMD) execution
model, where devices are organized into a multi-dimensional \emph{device mesh}~\cite{torch_dcp,zhao2023pytorch,megascaledata}.
At each training step $s$, the distributed optimizer consumes a \emph{Global Batch}
$\mathcal{B}_s = \{x_1, x_2, \ldots, x_N\}$, a finite set of $N$ training samples
used to compute the gradient update:
\begin{equation}
  \theta_{s+1} \leftarrow \theta_s - \eta\,\nabla\mathcal{L}(\mathcal{B}_s;\,\theta_s).
\end{equation}
This batch is partitioned across the $D \times C$ data-relevant positions in the
device mesh. Data Parallelism (DP) requires each replica to consume an independent
subset of samples. Context Parallelism (CP) splits each sample's token sequence
across ranks within a replica, so CP ranks share the same samples but consume
different token chunks. Tensor Parallelism (TP) and Pipeline Parallelism (PP)
partition model parameters rather than input data, so ranks within the same TP or
PP group receive identical input. The data plane therefore needs only to distinguish
$D \times C$ positions in the mesh, where $D$ is the DP world size and $C$ is the
CP degree; TP and PP are transparent to data delivery.

This structure imposes two hard requirements on the data plane. First,
\textit{intra-batch consistency}: all ranks must derive their projections from the
same $\mathcal{B}_s$. The batch must become atomically visible across the cluster
before any rank begins its optimization step. Partial visibility, where some ranks
see a complete batch while others see only a prefix, leads to inconsistent gradient
aggregation and parameter corruption. Second, \textit{inter-batch ordering}: the
sequence $(\mathcal{B}_1, \mathcal{B}_2, \ldots)$ must be strictly monotonic and
globally agreed upon. If different ranks consume batches in different orders, their
local model states diverge immediately.

Checkpoint recovery tightens these requirements further. Periodic distributed
checkpoints persist the model weights alongside the data-plane state. Upon recovery,
training must resume from the exact batch where it left off, consuming the same
sequence as in the original run. The batch sequence is therefore not merely a
delivery artifact but a durable, replayable history that must be preserved
independently of the training process.

\stitle{Runtime-dependent preprocessing.}
The samples that constitute a Global Batch do not arrive ready to consume. In
modern LFM training, raw input passes through a multi-stage preprocessing pipeline
before it can be packed into a batch and delivered to the optimizer. This pipeline
is runtime-dependent: its compute cost, output volume, and latency are determined
by input content and the current training configuration, not by static metadata.

\begin{figure}[!t]
  \centering
  \includegraphics[width=\linewidth]{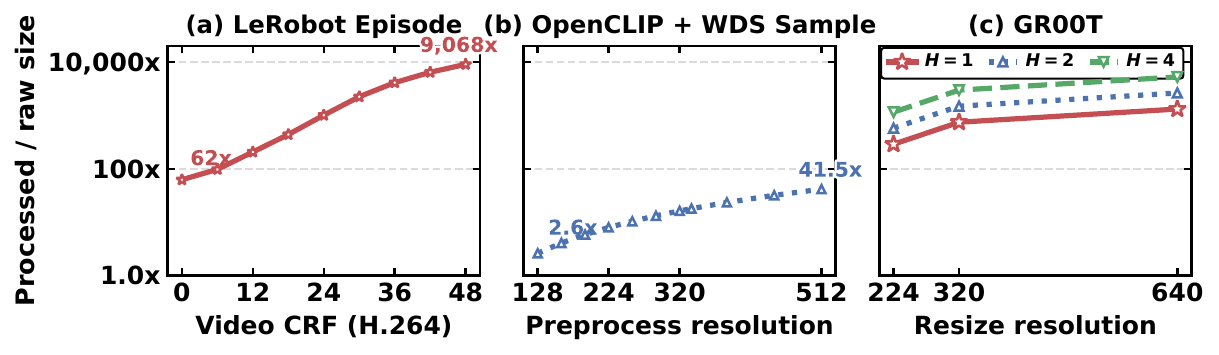}
  \caption{Training-time preprocessing inflates data volume by large,
  configuration-dependent factors. The expansion ratio ranges from
  $62\times$--$9{,}068\times$ for LeRobot episodes (varying H.264 CRF),
  $2.6\times$--$41.5\times$ for OpenCLIP samples (varying resolution), and
  $288\times$--$5{,}263\times$ in the GR00T path used in our experiments
  (varying resize resolution and observation history).}
  \label{fig:motivation:inflation}
\end{figure}

Multimodal training provides a representative example. A video-text sample requires
video decoding, frame extraction at a resolution set by the model's current input
specification, and spatial or temporal subsampling. The output volume depends on
video duration, codec, and target resolution, none of which can be bounded
statically. In a LeRobot~\cite{cadene2024lerobot} cloth-folding episode with three
$480\times640$ camera streams, the processed episode reaches $22.2$\,GiB while
the raw episode ranges from $364.7$\,MiB to $2.5$\,MiB as the video CRF changes,
implying $62\times$--$9{,}068\times$ expansion
(Figure~\ref{fig:motivation:inflation}a). In our GR00T end-to-end experiments,
replaying the dataset reader, sample transforms, Eagle batch transform, and
\texttt{Batch.to\_bytes()} serialization on three \texttt{adjust\_bottle} episodes
yields $288.3\times$--$5{,}263.1\times$ ready-to-train expansion as resize
resolution changes from $224$ to $640$ and observation history grows from $1$ to
$4$ (Figure~\ref{fig:motivation:inflation}c); jumps are discrete because visual
tokenization cost follows tile-count plateaus. Per-sample latency is also highly
heterogeneous: short and long clips can differ in processing time by orders of
magnitude, introducing stragglers that cannot be predicted or balanced statically.

A second example is online experience rollout in reinforcement learning from human
or verifiable feedback (RLHF/RLVR)~\cite{ouyang2022training,deepseek_r1,shao2024deepseekmath}.
In these workloads, the model itself generates training data by rolling out its
current policy; the volume and composition of each rollout batch depend on the model
state at that step and cannot be known in advance.

The same pattern holds for image-text training: an OpenCLIP~\cite{ilharco2021openclip}-style
WebDataset~\cite{aizman2020webdataset} sample of $74.0$\,KiB expands to
$192.6$\,KiB--$3.0$\,MiB as training resolution changes from $128$ to $512$, a
$2.6\times$--$41.5\times$ expansion (Figure~\ref{fig:motivation:inflation}b).

Across all three cases, preprocessing volume is large, configuration-dependent, and
unknown until execution. The data plane must accommodate bursty, dynamically sized
production decoupled from the synchronous demands of training ranks; static resource
allocation and pre-partitioning are infeasible.

\begin{figure}[!t]
  \centering
  \includegraphics[width=\columnwidth,height=0.4\columnwidth]{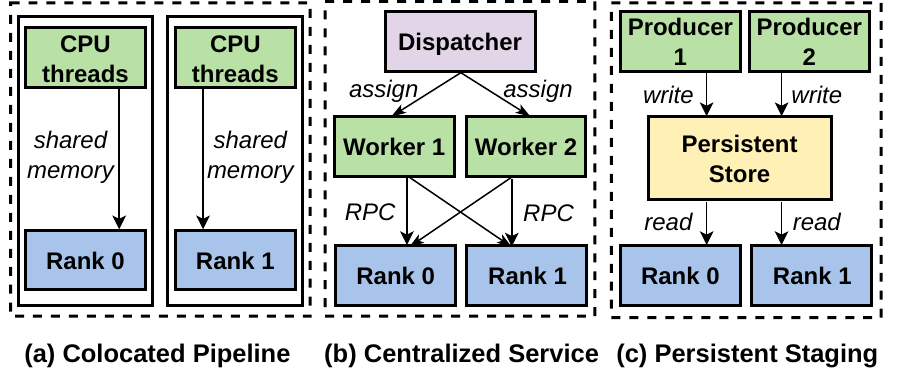}
  \caption{Three architectural patterns for training dataflow.
  Colocated dataloaders couple preprocessing with training on the same node;
  centralized services decouple them but deliver batches ephemerally via RPC;
  BatchWeave stages TGBs persistently on object storage, decoupling
  production from consumption while preserving batch-level semantics.}
  \label{fig:arch}
\end{figure}

\begin{figure}[!t]
  \centering
  \includegraphics[width=\columnwidth,height=0.6\columnwidth]{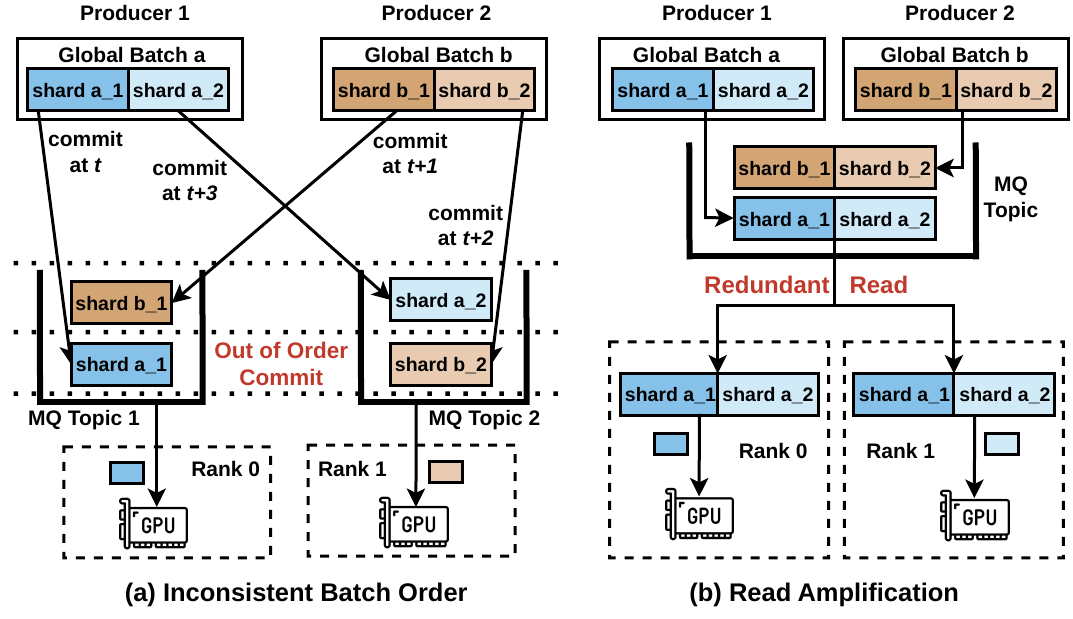}
  \caption{Two structural limitations of message queues for LFM training.
    \textbf{(a) Inconsistent Batch Order}: concurrent producers commit shards
    out of order, causing DP ranks to consume shards from different batches
    at the same step and corrupting gradient aggregation.
    \textbf{(b) Read Amplification}: delivering a full batch as one message
    forces every rank to download all $D$ shards, consuming only~$1/D$.}
  \label{fig:mq-limitations}
\end{figure}

\subsection{Pitfalls of Existing Solutions}
\label{sec:background:pitfalls}

Three architectural patterns have emerged for organizing the training dataflow.
All three fall short of the requirements identified in
\S\ref{sec:background:requirements} for LFM workloads.

\stitle{Colocated pipeline.}
In a colocated pipeline, preprocessing runs within the same process group as the
training ranks, typically on CPU threads or colocated nodes
(Figure~\ref{fig:arch}a). Representative systems include PyTorch
\texttt{DataLoader}~\cite{paszke2019pytorch} and tf.data~\cite{murray2021tfdata} in their default
configurations, as well as CoorDL~\cite{mohan2021coordl}, which extends this model
with coordinated cache sharing across colocated training jobs. This arrangement
provides natural semantic alignment: batch boundaries, ordering, and recovery are
managed within a single process group.

The fundamental limitation is the absence of failure isolation and elastic scaling.
A preprocessing failure stalls or terminates the entire training job, wasting the
full GPU allocation for the duration of recovery. Since preprocessing and training
share the same resource pool, the two cannot be scaled independently.
Furthermore, even with careful threading and pipelining, preprocessing threads
compete with the training process for CPU cycles and memory bandwidth on the same
node, imposing a throughput ceiling that cannot be lifted without architectural
decoupling.

\stitle{Disaggregated dataloaders with centralized services.}
To scale preprocessing independently, disaggregated systems move it to a separate
worker pool coordinated by a centralized dispatcher (Figure~\ref{fig:arch}b).
Representative systems include tf.data service~\cite{audibert2023tfdata_service},
Cachew~\cite{graur2022cachew}, FastFlow~\cite{um2023fastflow}, and
MegaScale-Data~\cite{megascaledata}. This decoupling improves GPU utilization and
allows preprocessing resources to scale independently.

However, these systems deliver batches ephemerally: once a batch is transferred to
the training rank via RPC, it is discarded. There is no persistent batch history,
no atomic visibility guarantee across ranks, and no mechanism for replaying a
specific batch after a failure or rollback. The centralized dispatcher is also a
single point of failure: its crash stalls the entire pipeline, and all coordination
traffic passing through it creates a scaling bottleneck.

\stitle{Disaggregated dataloaders with persistent staging.}
A third approach interposes a persistent storage layer between producers and
consumers (Figure~\ref{fig:arch}c). Producers write preprocessed data
to a shared store; consumers read from it independently. This design supports
failure isolation, elastic scaling, and data persistence. However, realizing
correct training semantics on top of such a layer requires solving problems that
existing persistent storage abstractions do not address: batch-level atomicity
across concurrent producers, globally consistent ordering without a central
coordinator, and checkpoint-aligned lifecycle management. These are the challenges
that BatchWeave is designed to solve, as we detail in \S\ref{sec:overview}.

General-purpose message queues such as Kafka are a natural candidate for the
persistent store, but their record abstraction is fundamentally misaligned with
training semantics. As Figure~\ref{fig:mq-limitations} illustrates, MQ systems
offer no batch-level primitive: a Global Batch is not an addressable entity but
an implicit collection of independent records. This mismatch takes two forms.
If each rank-specific shard is delivered as a separate message, out-of-order
commits from concurrent producers cause different DP ranks to assemble shards
from different batches at the same step, silently corrupting gradient
aggregation~(a). If the entire Global Batch is packed into one message, every
rank must download the full payload and discard all but its own shard, incurring
$D$-fold read amplification~(b). Imposing batch-level atomicity via a consumer-side
barrier or external coordinator would satisfy the ordering requirement but
reintroduces a centralized component with its own failure domain, which is
contrary to BatchWeave's storage-native design.

\subsection{Opportunities With Object Storage}
\label{sec:background:objstore}

Object storage systems such as Amazon S3, Google Cloud Storage, and Azure Blob
Storage exhibit three properties that make them a natural candidate for the
persistent staging layer described above. First, objects are written atomically and
immutably, which aligns with append-only data materialization. Second, access is
fully decentralized: any authorized producer or consumer can read or write without
routing through a dedicated coordination service, with no partitions to assign and no capacity to
pre-allocate. Third, a new training job requires only a fresh namespace prefix,
with no cold-start provisioning; storage capacity scales elastically with usage
without operator intervention.

Beyond these substrate properties, direct writes support elastic production:
producers join and exit without registration, a crash affects only its own
in-flight objects, and aggregate throughput scales with the producer pool.
Eliminating a central coordinator removes both a failure domain and a scaling
ceiling, and is structurally consistent with the SPMD execution model.

\section{BatchWeave Overview} \label{sec:overview}

Object storage provides durable, decentralized, immutable object writes, but
offers no mechanism for expressing relationships between objects, no way to define
what constitutes a complete batch, and no notion of training progress. Our key
insight is to adapt the versioned-manifest architecture of lakehouse
systems~\cite{delta,iceberg,lance}: dataset state is a monotonically growing sequence of
immutable manifest files, and successful publication of a new version is
serialized by a conditional write on the next version name.

Lakehouse manifests encode partition layouts for analytical query planning and are
written infrequently. Applying this pattern to a training data plane requires the
manifest to encode \emph{batch boundaries} rather than partition layouts, the
commit protocol to sustain high frequency under concurrent production, and the
system to durably track training-progress state (producer offsets and consumer watermarks)
so that fault recovery and lifecycle management are driven by checkpoint progress.

\begin{figure}[!t]
  \centering
  \includegraphics[width=\columnwidth]{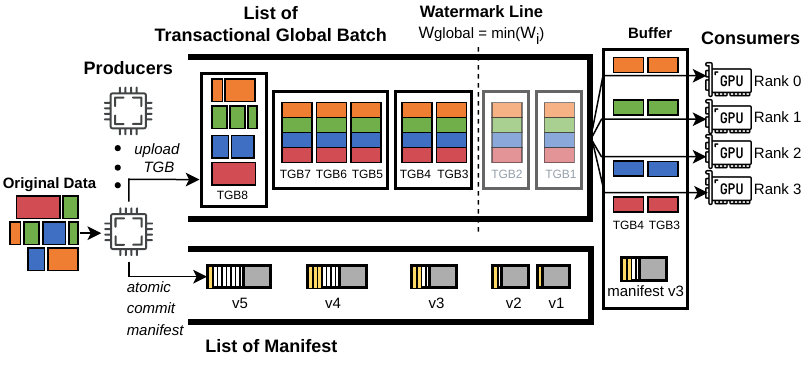}
  \caption{BatchWeave architecture. Producers write TGBs directly to object
    storage and expose them atomically via a versioned manifest; consumers issue
    per-rank range reads against the objects of committed TGBs. The global watermark
    $W_{\text{global}} = \min(W_i)$ bounds which TGBs are eligible for
    reclamation. The critical path consists of object writes, conditional
    manifest publication, and direct range reads.
  }
  \label{fig:overview}
\end{figure}

Figure~\ref{fig:overview} shows the overall architecture. BatchWeave is implemented
entirely as a client-side library with no server-side processes. It exposes two
lightweight clients: a \textbf{producer client}, embedded in preprocessing workers,
and a \textbf{consumer client}, embedded in each training rank. The
\textbf{object store} is the sole shared substrate; no dispatcher or dedicated
coordination service sits in the critical path.

The central abstraction is the \textbf{Transactional Global Batch (TGB)}, which
gives $\mathcal{B}_s$ a durable, first-class representation on object storage.
A TGB is a logically atomic batch unit materialized as one or more immutable
object-store objects. Its defining property is manifest-gated visibility:
writing the objects alone does not make the TGB visible; the TGB becomes visible
only when a committed manifest version records it and its position in the global
step sequence. A single manifest commit may publish one or more newly written
TGBs at once, so multiple TGBs can share the same first-visible manifest
version. This provides atomicity through manifest-gated visibility, durability
through object storage persistence, and globally consistent ordering through a
linearized manifest version sequence.

BatchWeave makes no assumptions about producer count or stability. Producers may
join or leave at any time, fail and restart, or produce data at varying rates,
and producer and consumer lifecycles need not be synchronized.

\subsection{Architecture and Data Flow}
\label{sec:overview:arch}

The data flow in Figure~\ref{fig:overview} proceeds in four stages.

\stitle{Stage 1: TGB materialization.} The producer client serializes
preprocessing output for one or more newly constructed TGBs into immutable
object-store objects. This step requires no coordination: multiple producer
clients write in parallel without communicating with each other or with
consumer clients.

\stitle{Stage 2: Manifest commit.} Once a producer client has accumulated one
or more newly constructed TGBs, it commits them to the manifest sequence using a
conditional put: the producer writes a new manifest object named with the next
version number, succeeding only if no object with that name already exists. If
a concurrent producer has already committed the same version, the write fails
and the producer rebases onto the current manifest before retrying. The
Decentralized Adaptive Commit (DAC) algorithm governs when each producer
initiates a commit, balancing data freshness against conflict rate. A
successful commit atomically makes the corresponding TGBs visible to all
consumer clients.

\stitle{Stage 3: Batch consumption.} The consumer client polls the manifest for
new versions and fetches the rank-specific byte range from the objects of newly
visible TGBs. It computes its projection locally from a lightweight footer
index cached per TGB, requiring no inter-rank communication, and prefetches the
referenced objects asynchronously to hide object store read latency.

\stitle{Stage 4: Lifecycle management.} After each successful distributed
checkpoint, the training framework records the consumer's current position as a
watermark. BatchWeave derives a global safety boundary
$W_{\text{global}} = \min_i(W_i)$; TGBs and manifest versions below this
boundary are eligible for reclamation, tying data retention to checkpoint
progress.

Sections~\ref{sec:design} and~\ref{sec:control} describe the TGB data plane and
the manifest-based control plane in detail.

\section{Transactional Global Batch} \label{sec:design}

The Transactional Global Batch (TGB) abstraction gives each $\mathcal{B}_s$ a
durable, atomically visible representation on object storage. This section describes
how TGBs are physically laid out to support the N-dimensional parallelism of
modern LFM training, how the manifest represents their logical structure and
ordering, how the combination enforces atomic visibility, and how the consumer
client addresses rank-specific data without inter-rank coordination.

\subsection{TGB Layout}
\label{sec:design:layout}

Each TGB is materialized as one or more immutable objects written to the object
store by a producer client, containing the training-ready data for one Global
Batch $\mathcal{B}_s$. These objects are write-once: once persisted, their
content never changes, allowing producer clients to write independently and
consumer clients to cache aggressively without coherence overhead.

The internal layout follows the data-sharing structure of the device mesh.
As established in \S\ref{sec:background:requirements}, only DP and CP require
distinct data across ranks; TP and PP are transparent to data delivery.
Accordingly, the data within a TGB is organized into $D \times C$ contiguous
data slices, where slice $(d, c)$ contains the token chunk for CP rank $c$ of
DP replica $d$. Each TGB includes a lightweight footer index recording the byte
offset and length of every $(d, c)$ slice within its constituent objects. A
consumer client reads this footer once per TGB, caches it locally, and
thereafter fetches its data via targeted range reads.

As a concrete example, consider a mesh with $D=2$ DP replicas and $C=2$ CP ranks,
giving $4$ slices per TGB. Replica 0 reads slices $(0,0)$ and $(0,1)$
for its two CP ranks; replica 1 reads $(1,0)$ and $(1,1)$. Any TP or PP ranks
within the same DP replica and CP group derive identical $(d,c)$ coordinates and
read the same slice. A 16-GPU job using $D=2$, $C=2$, $\text{TP}=2$, $\text{PP}=2$
therefore resolves exactly 4 distinct $(d,c)$ slices per TGB, one per $(d,c)$ pair,
regardless of the TP and PP degree.

This layout eliminates the read amplification of centralized disaggregated systems,
where each rank receives the full batch and discards the portions belonging to other
ranks, incurring $D \times C$-fold amplification. In BatchWeave, each rank reads
only its own data slice regardless of parallelism degree.

\stitle{Topology reconfiguration.}
Multi-stage LFM pipelines may use different parallelism degrees across stages, and
practitioners frequently adjust TP, PP, or DP when resuming from a checkpoint.
Rather than rewriting materialized TGBs, the consumer client remaps its position
in the new device mesh locally. Changes to TP or PP degree leave data distribution
unchanged; the consumer simply recomputes its $(d, c)$ coordinates. Changes to DP
or CP degree alter how many data slices are needed per step: when DP world size
doubles, the consumer reads two consecutive TGBs as one logical step; when it
halves, the consumer reads alternating data slices from a single TGB across two
steps. CP reconfiguration follows the same logic along the token-chunk dimension.
Remapping requires no server-side intervention, no data rewrite, and no
coordination with other ranks.

\subsection{Manifest Structure}
\label{sec:design:manifest}

The manifest is the logical control structure of BatchWeave. It is a versioned
sequence of immutable metadata objects stored on the object store, where each
version captures the complete state of the dataset at a point in time. A manifest
version $M_v$ contains two primary components.

The \textbf{TGB list} is an ordered sequence of TGB descriptors, each
recording which step index the TGB corresponds to together with the object-store
locations and layout metadata needed to read it. The TGB list defines the
authoritative step sequence: the
$s$-th step $\mathcal{B}_s$ is the set of samples identified by the $s$-th entry
in the list, regardless of when or by whom the corresponding TGB was written.

The \textbf{version sequence} consists of immutable manifest objects named by their
version number (e.g., \texttt{00000011.manifest}). A producer commits by writing the
next object under a conditional put that succeeds only if the name is unclaimed;
conflict causes immediate failure with no further effect. This single write
atomically advances the version and makes new TGBs visible, with no separate
pointer to update. Readers follow progress by probing for higher-numbered manifest
objects.

In addition to these two components, BatchWeave maintains a \textbf{per-producer state map} that records the stream offset up to which each
producer has successfully committed. It allows a producer that lost a commit race
to determine, after rebasing, which of its TGBs are already incorporated.

\subsection{Atomic Visibility}
\label{sec:design:visibility}

A TGB is opaque to consumer clients until it is referenced in a committed
manifest version. Consumer clients read only the TGB list from the current
manifest version and fetch only the objects referenced by those TGB
descriptors; they never scan the object store directly.
Because a manifest version is a single atomically written object, every version a
consumer observes is internally consistent: it references a complete, closed set
of TGBs for each step it exposes.

If a producer fails mid-commit, the version pointer is not updated and any
partially written TGB data remains invisible. The next successful commit will
include or exclude that data in a consistent way, as governed by the rebase
protocol in \S\ref{sec:control:commit}.

\subsection{Consumer Cursor and Deterministic Projection}
\label{sec:design:cursor}

Each consumer client maintains a cursor $\langle V, S \rangle$, where $V$
identifies the manifest version being read and $S$ the step index within that
version's TGB list. Given this cursor, the consumer independently computes its
data slice: it derives its $(d, c)$ coordinates from environment variables
(e.g., \texttt{RANK}, \texttt{WORLD\_SIZE}, and the configured DP and CP degrees),
looks up the byte offset and length for data slice $(d, c)$ in the cached footer
index, and issues targeted range reads against the corresponding object-store
objects. Ranks within the same TP or PP group
derive identical $(d, c)$ coordinates and read the same data slice without
inter-rank communication; DP replicas derive disjoint coordinates and read disjoint
data from the same TGB, keeping gradient aggregation mathematically consistent.

After consuming step $S$, the consumer increments $S$; when $S$ reaches the end
of the current TGB list, it polls for a new manifest version and updates $V$.
This polling is the only point at which the consumer touches the manifest; all
data reads go directly to the object store from addresses resolved through the
footer index.

The cursor is the recovery interface between BatchWeave and the training framework:
after each successful distributed checkpoint, the framework persists the current
cursor alongside the model weights. The persistence protocol and its recovery
semantics are described in \S\ref{sec:control:fault}.

\section{Manifest-Based Control Plane} \label{sec:control}

The manifest-based control plane of BatchWeave manages three responsibilities:
linearizing the global batch sequence across concurrent producers, regulating commit cadence
to sustain throughput as the manifest grows, and ensuring fault tolerance and
correct lifecycle management under producer failover and consumer restart. All three
are realized without a central coordinator, using only object storage primitives
and durable per-producer state.

\subsection{Commit and Rebase Protocol}
\label{sec:control:commit}

Producer clients commit TGBs to the manifest using a conditional put protocol
that serializes concurrent updates through object storage. The protocol proceeds in three
steps. First, the producer starts from its current local manifest view $M_v$.
Second, it constructs a candidate manifest $M_{v+1}$ by appending its local TGB
references to the TGB list of $M_v$ together with updated producer metadata.
Third, it attempts to write $M_{v+1}$ as
\texttt{(v+1).manifest} via conditional put, succeeding only if the name is
unclaimed.

On conflict, the producer fetches the committed winner manifest $M_{v'}$ and
rebases onto it: it appends its local TGB references to the winner's list and
updates its own producer metadata, then retries. A producer need not observe the
latest manifest before each attempt: starting from a stale base can only cause
additional failed conditional writes, while rebasing on the winner preserves
correctness.

Because the conditional put is atomic, no two producers can commit the same version
number, and the append-only union merge ensures no committed TGB is ever lost
during a rebase. Version numbers are strictly monotonically increasing and never
reused, so there is no ABA hazard: a producer that reads version $v$ and later
finds version $v'$ in a conflict can safely take $M_{v'}$ as the rebase base,
knowing its own TGBs are either already present (verifiable via persisted producer state) or must
be appended. The result is a linearized step history consistent across all
data-parallel ranks, with no inter-producer communication.

\subsection{Decentralized Adaptive Commit (DAC)}
\label{sec:control:dac}

Sustaining high producer ingestion throughput requires careful regulation of commit
cadence. The dominant cost per commit cycle is manifest I/O, and this cost grows
monotonically as the manifest accumulates entries. A fixed commit interval becomes
too aggressive over time, wasting an increasing fraction of producer time on failed
commits. The central observation driving DAC is that manifest I/O time, the
\emph{fragile window} $\tau_v$, is directly measurable at runtime and subsumes
payload size, producer count, and network conditions into a single online-observable
quantity. Adapting the commit interval to $\hat{\tau}_v$ achieves stable throughput
without static tuning.

Reactive heuristics such as AIMD~\cite{jacobson1988congestion} adjust the commit
interval based on observed outcomes but provide no quantitative guarantee on wasted
time. DAC maintains an explicit overhead budget $\delta$ and conflict budget
$\varepsilon$, deriving a closed-form lower bound on $T$ from each constraint
(Equations~\ref{eq:tconf}--\ref{eq:tcost}) and taking their maximum. Because
$\tau_v$ directly captures manifest I/O cost, DAC adapts to manifest growth
automatically with no separate tracking mechanism.

Each producer client controls a post-attempt waiting gap $T$: after any commit
attempt, whether it succeeds or fails, the producer waits $T$ before initiating the
next attempt. A commit proceeds by reading the current manifest version,
constructing a new manifest object named with the next version number, and
submitting it via conditional put. The fragile window $\tau_v$ is the interval from
reading the current version to completing the write attempt: if another producer
successfully commits the same version number within this window, the current commit
fails. Under this control law, successive attempt starts are separated by
approximately $T + \tau_v$.

\stitle{Probabilistic model.}
We model the attempt starts of the other $N-1$ producers as independent Poisson
processes with rate $1 / (T + \tau_v)$. Under this renewal approximation, the
probability that at least one competing attempt enters the fragile window of length
$\tau_v$ is:

\begin{equation}
 p_{\text{conflict}}(T) = 1 - e^{-(N-1)\tau_v / (T + \tau_v)}
\label{eq:conflict}
\end{equation}

\stitle{Attempt-duty budget.}
Each attempt cycle spends $\tau_v$ seconds on manifest I/O and lasts
approximately $T + \tau_v$ seconds, so the resulting commit-duty factor is:

\begin{equation}
d(T) = \frac{\tau_v}{T + \tau_v}
\label{eq:duty}
\end{equation}

DAC enforces two explicit budgets: a conflict budget $\varepsilon$ on
$p_{\text{conflict}}(T)$ and a duty budget $\delta$ on $d(T)$.
These define the feasible set
\begin{equation}
\mathcal{F} = \{T \geq 0 \;:\; p_{\text{conflict}}(T) \leq \varepsilon,\;
d(T) \leq \delta\}
\label{eq:feasible}
\end{equation}
and DAC chooses the smallest feasible gap,
\begin{equation}
T^* = \inf \mathcal{F}
\label{eq:objective}
\end{equation}
which maximizes freshness while respecting both budgets. Both
$p_{\text{conflict}}(T)$ and $d(T)$ decrease monotonically with $T$, so $T^*$ is
obtained by taking the maximum of the two resulting lower bounds.

\stitle{Online algorithm.}
In practice, $\tau_v$ is estimated online via an exponential moving average, and
$N$ is read dynamically from committed producer state after each rebase, allowing
DAC to track changes in the producer pool without inter-producer communication:
\begin{equation}
\hat{\tau}_v \leftarrow (1 - \alpha)\,\hat{\tau}_v + \alpha\,\tau_v^{\text{obs}}
\end{equation}

Rather than solving Equation~\ref{eq:objective} numerically, DAC computes the
required gap directly from the two constraints. Solving
$p_{\text{conflict}}(T) \leq \varepsilon$ yields:
\begin{equation}
T \geq T_{\text{conf}} =
\max\!\left(0,\; \frac{(N-1)\,\hat{\tau}_v}{-\ln(1 - \varepsilon)} - \hat{\tau}_v\right)
\label{eq:tconf}
\end{equation}
and solving $d(T) \leq \delta$ yields:
\begin{equation}
T \geq T_{\text{cost}} = \frac{1 - \delta}{\delta}\,\hat{\tau}_v
\label{eq:tcost}
\end{equation}
The minimal feasible gap is therefore
\begin{equation}
T^* = \max(T_{\text{conf}},\; T_{\text{cost}})
\label{eq:tstar}
\end{equation}
and the target gap is a jittered version of this optimum used to desynchronize
producers:
\begin{equation}
\text{gap} = T^* \cdot (1 + \rho \cdot U)
\label{eq:gap}
\end{equation}
where $U \sim \text{Uniform}(0, 1)$ and $\rho$ is the jitter magnitude. This gap
is recomputed after every commit attempt and takes effect immediately for the next
attempt. When bursts of correlated commits occur, the EMA update raises
$\hat{\tau}_v$ and widens the gap on the next cycle, self-correcting within a few
attempt periods.

On a successful commit, the producer clears its local data buffer. On a failed
commit, the producer rebases onto the current manifest, merging its local TGB
references and updating $N$, before retrying. Algorithm~\ref{alg:dac}
summarizes the full procedure, including the finalization phase in which the
producer drains remaining uncommitted data before exiting.

\begin{algorithm}[t]
\caption{Decentralized Adaptive Commit (DAC)}
\label{alg:dac}
\begin{algorithmic}[1]
\State \textbf{Parameters:} $\delta$ (overhead budget), $\varepsilon$ (conflict
       budget), $\alpha$ (EMA coefficient), $\rho$ (jitter magnitude)
\State \textbf{Initialize:} $\hat{\tau}_v \leftarrow 0$,\ $\text{gap} \leftarrow 0$,\
       $N \leftarrow 1$,\ $t_{\text{last}} \leftarrow \textsc{Now}()$
\Loop
    \State $\textsc{WriteTGB}()$ \Comment{Materialize data; no coordination needed}
    \If{$\textsc{Now}() - t_{\text{last}} \geq \text{gap}$}
        \State $t_0 \leftarrow \textsc{Now}()$
        \State $\text{success},\ M \leftarrow \textsc{TryCommit}()$
        \State $\tau_v^{\text{obs}} \leftarrow \textsc{Now}() - t_0$
        \State $\hat{\tau}_v \leftarrow (1-\alpha)\,\hat{\tau}_v + \alpha\,\tau_v^{\text{obs}}$
        \Comment{Update EMA regardless of outcome}
        \If{$\text{success}$}
            \State $\textsc{ClearBuffer}()$
        \Else
            \State $M \leftarrow \textsc{Rebase}()$ \Comment{Merge onto current manifest}
        \EndIf
        \State $N \leftarrow |\textsc{ProducerStateMap}(M)|$
        \Comment{Dynamic producer count}
        \State $T_{\text{conf}} \leftarrow \max(0,\; (N-1)\,\hat{\tau}_v \;/\; {-\ln(1-\varepsilon)} - \hat{\tau}_v)$
        \State $T_{\text{cost}} \leftarrow (1-\delta)\,\hat{\tau}_v \;/\; \delta$
        \State $\text{gap}_{\text{new}} \leftarrow \max(T_{\text{conf}},\; T_{\text{cost}}) \cdot (1 + \rho \cdot \textsc{Uniform}(0,1))$
        \State $\text{gap} \leftarrow \text{gap}_{\text{new}}$
        \State $t_{\text{last}} \leftarrow \textsc{Now}()$
    \EndIf
\EndLoop
\end{algorithmic}
\end{algorithm}

\subsection{Fault Tolerance and Lifecycle Management}
\label{sec:control:fault}

BatchWeave provides end-to-end exactly-once semantics across the full pipeline.
The producer and consumer sides use different mechanisms, each suited to the
structure of the state they must preserve. A single shared primitive, the watermark,
ties consumer fault tolerance to lifecycle management, serving both purposes without
additional infrastructure.

\stitle{Producer fault tolerance.}
Producer resumption state (the stream offset up to which TGBs have been
successfully committed) lives in process-local memory and is lost on failure.
Without a recovery mechanism, a replacement process must restart from a
conservative earlier position, re-producing and re-submitting data and yielding
at-least-once semantics.

BatchWeave avoids this by durably persisting producer resumption state as part
of the commit protocol. Each producer is assigned a stable \texttt{producer\_id}
that persists across restarts. The resumption state is updated in lockstep with
committed TGBs, ensuring the persisted offset always reflects the last visible
TGB. A replacement process recovers the persisted state for its
\texttt{producer\_id} and resumes from the highest recorded offset with no
coordination with other producers or consumers.

Exactly-once delivery follows from two invariants: the persisted resumption state
is always consistent with the committed TGB list, and the conditional write
prevents two processes sharing a \texttt{producer\_id} from advancing state
concurrently.

\stitle{End-to-end exactly-once argument.}
The two mechanisms are decoupled and jointly sufficient: neither a producer
restart nor a consumer rollback disturbs the other side's state. The only 
required invariant is that all data up to the rolled-back step remains
available in the manifest, which is guaranteed by the watermark retention policy.

\stitle{Consumer fault tolerance and lifecycle management.}
After the training framework successfully writes a distributed checkpoint, each
consumer records its current manifest version $V$ as a \emph{watermark} persisted
alongside the model weights. On rollback, the consumer restores its cursor from the
checkpoint and resumes from the corresponding global batch with no data skipped and
no batch consumed twice.

The same watermarks drive lifecycle management. Existing streaming systems govern
retention by time or capacity thresholds, with no awareness of checkpoint state;
acknowledgment-based systems such as Pulsar treat delivery as a sufficient deletion
signal, but a training rank that has consumed a batch may not yet have checkpointed
the resulting model state, so a rollback can require replaying already-acknowledged
data. BatchWeave instead ties retention directly to checkpoint progress via:

\begin{equation}
W_{\text{global}} = \min_{i \in \{1,\ldots,N\}} W_i
\label{eq:watermark}
\end{equation}

Any manifest version $v < W_{\text{global}}$ and its associated TGB objects are
unreachable from any live checkpoint and eligible for reclamation. A background
process periodically computes $W_{\text{global}}$ from the latest checkpoint
watermarks and issues deletion requests accordingly. This process is outside the
critical path: a failure delays reclamation but does not affect training correctness
or throughput. Because object store deletions are idempotent and TGB objects are
immutable, it can be restarted at any time without coordination.

This design provides two guarantees. First, rollback safety: TGB objects are retained
until $W_{\text{global}}$ advances past them, so any live checkpoint always finds
the data it needs. Second, principled reclamation: as training progresses,
$W_{\text{global}}$ advances and the retained window stays proportional to the
checkpointing interval rather than the total training duration.

\section{Implementation} \label{sec:impl}
BatchWeave is implemented as a Python SDK with a Rust core. The Rust layer handles
latency-critical execution: multi-threaded object prefetching, manifest polling,
and the optimistic commit protocol. It is exposed to Python via
\texttt{PyO3}-based foreign function interfaces, allowing users to implement custom
preprocessing logic without recompilation. This split keeps systems-level
throughput and memory safety alongside the rapid iteration cycles typical of
LFM development.

BatchWeave uses \texttt{If-None-Match} semantics
(available on S3, GCS, Azure Blob Storage, and BOS~\cite{mantle}) to implement the
manifest commit protocol described in \S\ref{sec:control:commit}. Ordinary object
reads are used only to fetch committed manifests during rebase and catch-up;
observing a stale manifest can increase retries but cannot violate correctness.
No other storage-side coordination is required.

For the physical data layer, BatchWeave materializes each TGB as one or more
immutable objects on the object store. Our current prototype builds on an
internal fork of Lance~\cite{lance}, extended to support the additional metadata and
atomicity guarantees required by BatchWeave's commit protocol.

Because BatchWeave is a pure client-side library, it integrates with any training
framework that can invoke Python: a PyTorch training loop and a TensorFlow pipeline
can share the same BatchWeave namespace without modification, and switching
frameworks requires only replacing the consumer call site.

\section{Evaluation} \label{sec:eval}

We evaluate BatchWeave on four questions aligned with its core claims:
end-to-end training throughput, producer scalability and DAC stability,
consumer efficiency under TGB-aware layout, and checkpoint-driven storage
reclamation. We also measure the cost of exactly-once producer state persistence.

\subsection{Experimental Setup}
\label{sec:eval:setup}

\stitle{Infrastructure.}
All experiments run on a shared kjob cluster backed by Baidu Object
Storage (BOS)~\cite{mantle}. Each node has 64 CPU cores. Training uses 8 GPUs
and 8 trainer ranks per node; producer and consumer microbenchmarks use
CPU-only nodes. The largest end-to-end run uses 8 trainer nodes (64 ranks) on
NVIDIA H200 GPUs. Unless otherwise noted, the logical consumer world size in an
end-to-end run equals the number of trainer nodes times 8.

\stitle{Systems compared.}
\textbf{BatchWeave} is our full object-store-native system.
\textbf{Kafka} is a centralized queue baseline using a dedicated Baidu Cloud Kafka
cluster (16-core, 64\,GB-memory instances, model kafka.ga2.c16m64, three
replicas, 10\,TB cloud disk per node) provisioned exclusively per experiment
run; no two runs share a cluster. Kafka always uses strict TGB semantics: one
message carries exactly one complete TGB, and producer count matches
BatchWeave. This is the only deployment mode for Kafka that satisfies our
intra-batch consistency and inter-batch ordering requirements (\S\ref{sec:background:requirements})
without reintroducing a centralized coordinator; we evaluate it to characterize the
structural cost of centralized queue delivery under LFM training semantics.
For large-payload video workloads, the one-message-per-TGB constraint triggers
two failure modes: TGB payloads exceeding per-message byte limits, and
produce-request timeouts under peak queue-service load. We tuned all available parameters (\texttt{message.max.bytes},
\texttt{request.timeout.ms}, \texttt{delivery.timeout.ms}) to minimize failures,
but could not eliminate them entirely for the Qwen3-VL configurations.
\textbf{Local} is the expert-tuned colocated pipeline used in our
production training stack. Each trainer rank launches 12 local worker threads
for sample-level preprocessing, feeds transformed samples through a bounded
sample queue into a dedicated collator thread for sequence packing and batch
construction, and applies periodic garbage collection and cache cleanup to
stabilize memory usage over long runs. This represents the engineering limit of
the colocated architecture: further increasing worker count intensifies CPU and
memory bandwidth contention with the GPU training process on the same node, and
the preprocessing and training lifecycles remain coupled within the same process
group.

\stitle{Commit policy baselines.}
Producer microbenchmarks compare five commit-policy baselines:
\textbf{Naive} (commit every TGB), \textbf{FIXED10} and \textbf{FIXED100}
(commit every 10 or 100 TGBs), \textbf{INCR} (start at 10 and increase by one
on each conflict), and \textbf{AIMD} (Additive Increase Multiplicative Decrease, the classic
TCP-style congestion control policy~\cite{jacobson1988congestion}: increase the interval by a fixed addend on
success, halve it on conflict). Consumer microbenchmarks additionally compare
\textbf{dense-read}, which reads the full TGB object span and filters locally.

\stitle{Workloads.}
We use four workload families: end-to-end GR00T~\cite{groot} training,
end-to-end HoloAssist~\cite{wang2023holoassist} video SFT, end-to-end
BEHAVIOR-1K~\cite{li2023behavior1k} VLA training, and controlled data-plane
microbenchmarks.
All GR00T runs use a trainer-side DataLoader with \texttt{num\_workers=1} and
\texttt{prefetch\_factor=4}. HoloAssist runs train Qwen3-VL-30B-A3B~\cite{qwen3vl2025}
on the HoloAssist reasoning split, with online video decode, frame sampling at
2\,FPS, and 8--16 frames per sample. BEHAVIOR-1K runs train the
Qwen3-VL-30B-A3B-AE-0.5B VLA model on multi-camera robot demonstrations stored in
LeRobot~\cite{cadene2024lerobot} format, with online video decode, image
augmentation, state/action windowing, and multimodal sequence packing. Payload
sizes are 100\,KB, 1000\,KB, and 10000\,KB. Producer experiments sweep
8--128 producers at logical world size~32, with a 300\,s warmup and a
5-hour measurement window. The DAC commit-policy ablation fixes the producer
count at 32 to isolate the effect of manifest growth over the same duration. Consumer
experiments sweep logical world sizes 8\,/\,32\,/\,128 for 1{,}800\,s per
point. Consumer baselines always read from pre-materialized committed datasets
so that all strategies observe identical input. The exactly-once microbenchmark
sweeps payload sizes 100\,KB, 1000\,KB, and 10000\,KB with TGB sizes
8\,/\,32\,/\,128. It alternates producer-state metadata with a dummy-metadata
control on paired append inputs for one hour. Each operation commits one
TGB, intentionally stressing per-commit metadata overhead; normal
DAC-driven runs amortize this cost by committing larger batches.

\stitle{Metrics.}
End-to-end experiments report throughput (steps/s) and per-step latency over
time. Producer experiments report aggregate ingestion throughput, commit
success rate, and commit latency. Consumer experiments report effective
per-rank throughput, P50/P95 read latency, and read amplification. Lifecycle
experiments report object-store bytes over time.

\stitle{Methodology.}
Producer benchmarks exclude warmup; consumer benchmarks aggregate per-rank
structured traces. DAC uses a conflict budget of $\varepsilon = 0.05$ in
producer microbenchmarks and $\varepsilon = 0.20$ in end-to-end runs.
End-to-end runs start producers and trainers together rather than from a
pre-filled backlog, so reported step timing begins at first-batch arrival and
excludes only the initial producer warm-up. We otherwise use all common
recorded steps for each accepted run, without tail trimming. Omitted Kafka
points indicate no usable strict-TGB run at that configuration.

\subsection{End-to-End Training Performance}
\label{sec:eval:e2e}

\begin{figure*}[t]
  \centering
  \includegraphics[width=\textwidth]{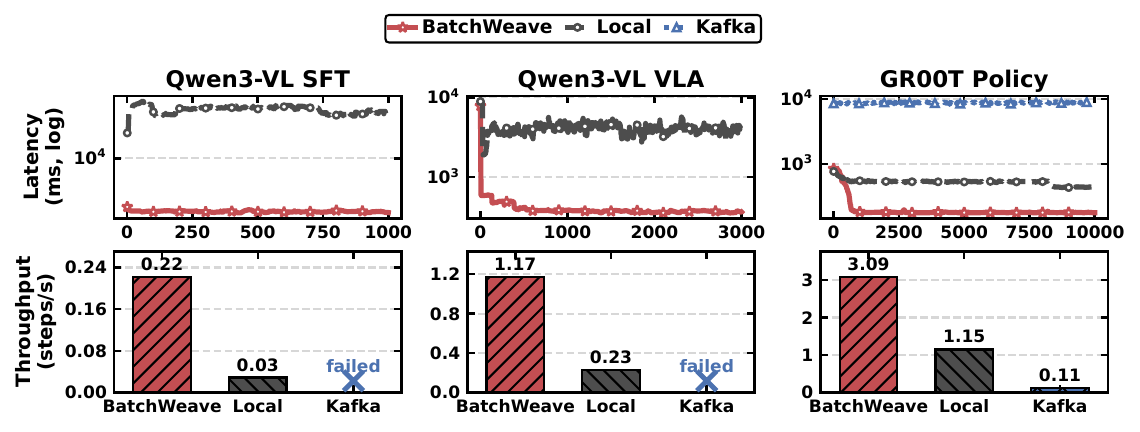}
  \caption{Per-step latency (ms, log) and end-to-end throughput (step/s) across three workloads.
  Kafka failed on the Qwen3-VL workloads (no usable strict-TGB run at 8 nodes);
  for GR00T it is shown over the longest measured prefix (9{,}694 steps), after
  which sustained queue-service backpressure caused per-step wait times to exceed an
  acceptable threshold.
  BatchWeave outperforms Local by 2.68--7.73$\times$ across all three workloads.}
  \Description{A two-row by three-column end-to-end training figure. Columns
  correspond to Qwen3-VL SFT, Qwen3-VL VLA, and GR00T Policy workloads. The top
  row plots per-step latency over training steps, and the bottom row reports
  throughput bars for BatchWeave, Local, and Kafka.}
  \label{fig:eval:e2e-main}
\end{figure*}

Unless otherwise noted, BatchWeave uses dedicated CPU producer nodes and no
train-side producers: 32 nodes for the Qwen3-VL workloads and 16 for GR00T.
Local uses the optimized in-rank threaded pipeline described above. Kafka uses a
separate producer job on CPU nodes of the same type, with the same producer count
as BatchWeave.

The bottom row of Figure~\ref{fig:eval:e2e-main} shows that BatchWeave sustains
3.09\,steps/s versus 1.15\,steps/s for Local and 0.11\,steps/s for Kafka: a
2.68$\times$ gain over Local and 27.3$\times$ over Kafka. The all-rank
step-latency P50/P95 values are 172/367\,ms for BatchWeave, 457/4{,}113\,ms for
Local, and 8{,}811/9{,}534\,ms for Kafka. The gain over Kafka follows from
publishing TGBs through direct object-store writes and conditional manifests. The gain over Local reflects structural
resource contention that in-rank engineering cannot eliminate: preprocessing
threads share CPU cores and memory bandwidth with the GPU training process on the
same node. Under GR00T's heavy preprocessing expansion
(\S\ref{sec:background:requirements}), the batch queue periodically empties and
the training rank stalls, producing the observed P95 of 4{,}113\,ms, nearly
10$\times$ the P50 of 457\,ms. BatchWeave eliminates this contention by running
preprocessing on dedicated nodes, yielding a P95 of 367\,ms.

\stitle{HoloAssist video SFT.}
BatchWeave sustains 0.222\,steps/s versus 0.029\,steps/s for Local on the
HoloAssist~\cite{wang2023holoassist} reasoning split (7.73$\times$ gain). The
all-rank P50/P95 latencies are 2.60/2.79\,s for BatchWeave and 33.7/42.7\,s for
Local. The gain reflects the same structural contention as GR00T: video decode,
frame sampling, and multimodal packing compete with training on the same nodes in
the local baseline, whereas BatchWeave offloads these to dedicated producer nodes.

\stitle{BEHAVIOR-1K VLA.}
BatchWeave sustains 1.17\,steps/s versus 0.227\,steps/s for Local on
BEHAVIOR-1K~\cite{li2023behavior1k} demonstrations in
LeRobot~\cite{cadene2024lerobot} format (5.17$\times$ gain). The all-rank
P50/P95 latencies are 0.374/0.577\,s for BatchWeave and 3.83/7.96\,s for Local.
Normal steps are sub-second, with occasional fetch stalls when the trainer catches
up to producer output; even with these stalls, disaggregating preprocessing cuts
total training time by 80.7\% over the optimized local pipeline. The 32-node
producer allocation reflects the heavier per-sample cost of multi-camera video
decode and multimodal packing in the Qwen3-VL workloads compared with GR00T's
16-node allocation.

\subsection{Producer Scalability and DAC}
\label{sec:eval:producer}

\begin{figure*}[t]
  \centering
  \includegraphics[width=0.52\textwidth]{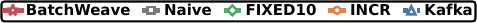}\\[-1pt]
  \includegraphics[width=\textwidth]{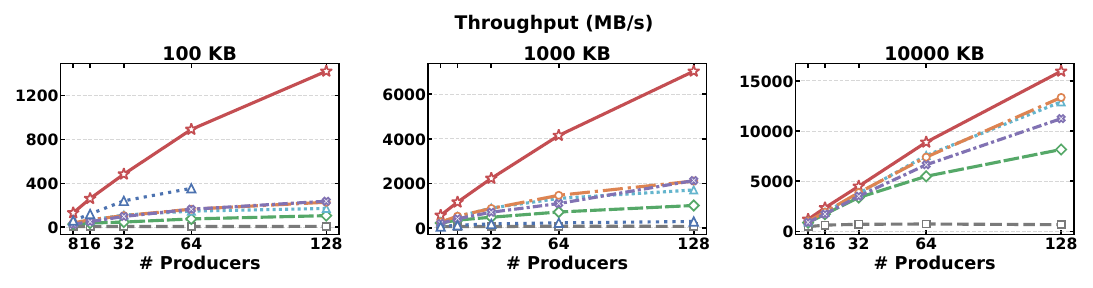}
  \caption{Producer ingestion throughput versus producer count across three payload sizes.
  BatchWeave is the only system that scales linearly with producer count;
  all baselines plateau or fail at high concurrency.
  Omitted Kafka points indicate no successful strict-TGB run at that configuration.}
  \label{fig:eval:producer-throughput}
\end{figure*}

\begin{figure}[t]
  \centering
  \includegraphics[width=\linewidth]{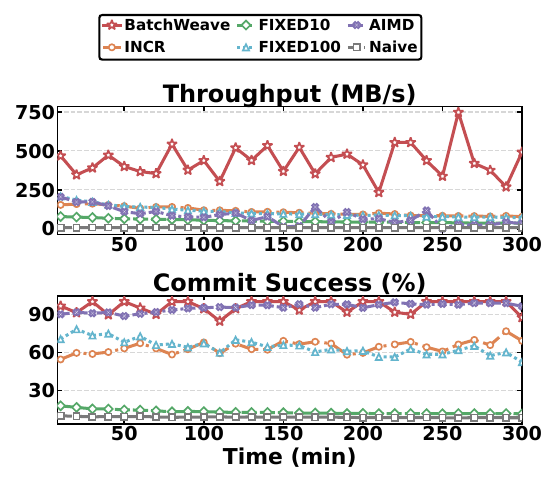}
  \caption{BatchWeave DAC-policy ablation with 32 producers over 5 hours.
  BatchWeave's DAC policy is the only policy that sustains both high throughput and commit success
  as the manifest grows; all fixed-interval and heuristic policies degrade.}
  \label{fig:eval:dac-ablation}
\end{figure}

\begin{figure}[t]
  \centering
  \includegraphics[width=\linewidth]{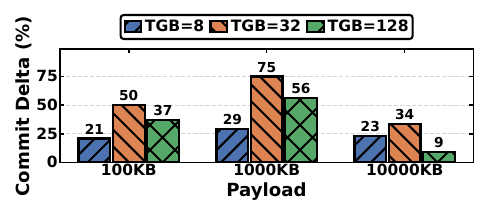}\\[2pt]
  \includegraphics[width=\linewidth]{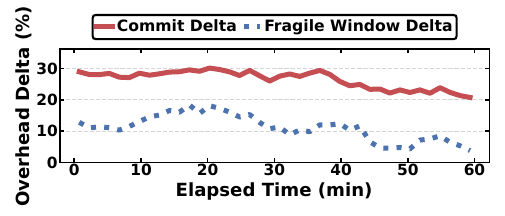}
  \caption{Overhead of exactly-once producer state persistence,
  measured against a dummy-metadata control.
  The per-commit cost (9.4\%--74.9\% at run start) declines over time because
  the fixed state size becomes negligible relative to growing commit I/O.}
  \label{fig:eval:eo-overhead}
\end{figure}

\begin{figure}[t]
  \centering
  \includegraphics[width=0.9\linewidth]{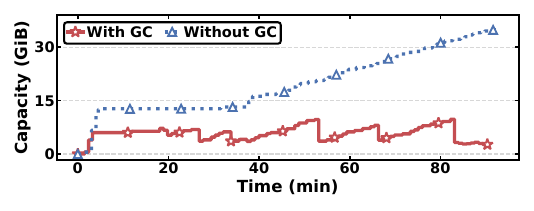}
  \caption{Checkpoint-driven storage reclamation over a 1{,}010-step run.
  With physical deletion enabled, peak capacity is capped at 9.76\,GiB
  (72.0\% reduction vs.\ 34.85\,GiB without deletion), confirming that
  checkpoint-written watermarks are a correct and tight reclamation signal.}
  \label{fig:eval:checkpoint}
\end{figure}

\stitle{Producer throughput scaling.}
Figure~\ref{fig:eval:producer-throughput} sweeps producer count and payload size,
comparing BatchWeave against Kafka and the commit-policy baselines. At 128
producers, BatchWeave reaches 1.42\,GB/s, 7.03\,GB/s, and 16.0\,GB/s at
100\,KB, 1000\,KB, and 10000\,KB. Against the strongest non-BatchWeave baseline
at the same producer count, this is 6.00$\times$, 3.31$\times$, and
1.20$\times$ faster, respectively. Kafka succeeds through 64 producers at
100\,KB and 128 producers at 1000\,KB, but no 10000\,KB strict-TGB point
succeeds. BatchWeave continues to scale, showing that direct object-store
publication is the dominant win. The narrower margin at 10000\,KB reflects a
shift in the dominant bottleneck: at large payloads, raw object-store write
bandwidth saturates for all strategies, leaving less room for commit-protocol
efficiency to differentiate them; at 100\,KB, per-request overhead and conflict
rate dominate, where direct publication and DAC provide the largest relative gain.

\stitle{DAC under manifest growth.}
Figure~\ref{fig:eval:dac-ablation} shows that DAC is the only policy that
sustains both success rate and throughput as the manifest grows over five
hours. With 32 producers on one CPU node, DAC averages
431.9\,MB/s at 96.3\% commit success. INCR reaches 110.5\,MB/s at 64.2\%,
FIXED100 107.5\,MB/s at 64.3\%, AIMD 79.0\,MB/s at 95.5\%, FIXED10
50.0\,MB/s at 13.1\%, and Naive 7.1\,MB/s at 9.0\%. As manifest growth raises
commit cost, policies that fail to widen the commit interval waste an
increasing fraction of time on retries. The measured conflict rate of DAC
stays close to the target $\varepsilon = 0.05$ in the producer
microbenchmarks, consistent with the Poisson model in practice.

\stitle{Overhead of exactly-once producer state.}
Figure~\ref{fig:eval:eo-overhead} isolates the per-commit cost of
durable producer state against a dummy-metadata control on paired append
inputs. Every TGB is committed immediately, intentionally stressing
per-commit metadata overhead rather than the amortized cost in normal DAC-driven
runs. The mean commit latency delta ranges from 9.4\% to 74.9\% across payload
and TGB sizes; the upper end occurs at TGB=8 and 100\,KB, where payload is
minimal and metadata overhead is not amortized. The bottom panel of Figure~\ref{fig:eval:eo-overhead} shows this relative cost declining
over the run: commit delta falls from 40.6\% to 32.4\% and fragile-window delta
from 50.5\% to 29.5\%, consistent with a fixed per-commit cost whose fraction
shrinks as per-commit work grows. In normal operation, DAC further amortizes
this cost by committing larger batches.

\subsection{Consumer Efficiency and Read Amplification}
\label{sec:eval:consumer}

\begin{figure}[t]
  \centering
  \includegraphics[width=\linewidth]{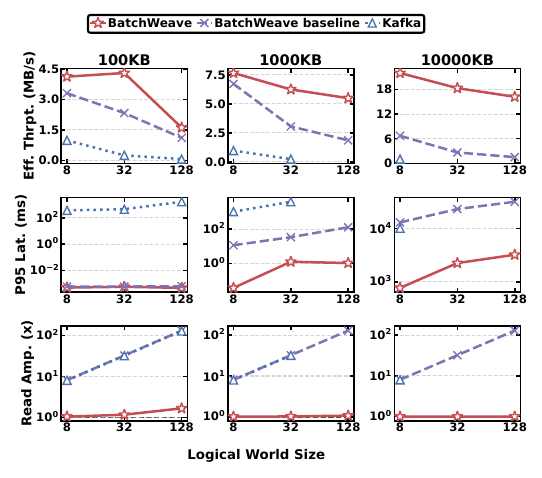}
  \caption{Consumer throughput, P95 latency, and read amplification across
  payload and world sizes.
  BatchWeave maintains near-$1\times$ read amplification at all scales because
  each rank issues a single range read targeting only its own data slice;
  dense-read and Kafka amplification grow linearly with world size.
  Omitted Kafka points indicate no successful strict-TGB run.}
  \label{fig:eval:consumer-sweep}
\end{figure}

Each rank issues a single range read targeting only its own data slice, so
BatchWeave's read amplification is near $1\times$ at all world sizes. The small
residual overhead comes from reading the footer index and object headers, a fixed
absolute cost whose fractional contribution shrinks with payload size. At 100\,KB
and world size~128 it accounts for the measured $1.67\times$; dense-read and Kafka
scale with world size because both deliver the full TGB before filtering. At 128
consumers and 100\,KB, BatchWeave reaches 1.61\,MB/s per rank at $1.67\times$
amplification, versus 1.11\,MB/s at $130.8\times$ for dense-read and 0.061\,MB/s
at $128.0\times$ for Kafka. At 10000\,KB and 128 consumers, BatchWeave cuts
P50/P95 latency from 121.6\,ms\,/\,31.4\,s to 22.8\,ms\,/\,3.21\,s relative to
dense-read, while improving per-rank throughput from 1.46\,MB/s to 16.2\,MB/s
(11.1$\times$). Kafka has no successful strict-TGB run at that configuration.
The P95 result is operationally significant: dense-read's 31.4\,s tail exceeds
typical per-step training budgets, so a straggler rank that falls behind on a
full-TGB read blocks the entire data-parallel group. BatchWeave's targeted
range reads keep the P95 below 4\,s across all measured configurations,
eliminating fetch stalls as a source of step-time variance.

\subsection{Checkpoint-Driven Lifecycle Management}
\label{sec:eval:lifecycle}

Figure~\ref{fig:eval:checkpoint} compares two otherwise
identical 1{,}010-step runs with checkpoints every 10 steps and
\texttt{max\_lag=80}: one with physical deletion enabled after logical trim,
and one without. The \texttt{max\_lag} parameter caps how many unacknowledged
TGBs producers may accumulate ahead of $W_{\text{global}}$, bounding peak
storage even if checkpointing temporarily stalls. Rank~0 samples total
object-store bytes every 60\,s and at each checkpoint boundary; the
resulting curve shows plateaus between samples and drops when an advancing
$W_{\text{global}}$ triggers reclamation at a checkpoint boundary.

Without physical deletion, capacity grows monotonically to 34.85\,GiB. With
physical deletion, it is capped at 9.76\,GiB, a 72.0\% reduction, confirming
that checkpoint-written watermarks are a correct control signal: data remains
available to live checkpoints and is reclaimed once no live checkpoint references
it.

\section{Related Work} \label{sec:related}

\subsection{Message Queues and Streaming Systems}
\label{sec:related:streaming}

Distributed message queues such as Apache Kafka~\cite{kreps2011kafka} and Apache
Pulsar~\cite{aguilar2015apache}, along with cloud-native variants including
WarpStream~\cite{warpstream2025}, AutoMQ~\cite{automq2024}, and
Ursa~\cite{merli2025ursa}, provide durable record delivery but retain the
record/offset abstraction. A training batch is not an addressable entity in these
systems; expressing batch-level atomicity requires an external coordination layer,
and retention is governed by time or capacity with no awareness of checkpoint state.
Stream processing frameworks such as Flink~\cite{carbone2015apache}, Spark
Structured Streaming~\cite{armbrust2018structured}, and
Dataflow~\cite{akidau2015dataflow} achieve fault tolerance through operator state
snapshots, but restoring operator state does not make consumed input sets
independently re-observable. BatchWeave externalizes the batch sequence as a
first-class persistent structure, making it replayable independently of any
processing logic.

\subsection{ML Training Data Pipelines}
\label{sec:related:ml}

Disaggregated systems such as tf.data service~\cite{audibert2023tfdata_service},
Cachew~\cite{graur2022cachew}, FastFlow~\cite{um2023fastflow},
FusionFlow~\cite{kim2023fusionflow}, and Cedar~\cite{zhao2024cedar} decouple
preprocessing from training to prevent GPU stalls, while CoorDL~\cite{mohan2021coordl}
and Pecan~\cite{graur2024pecan} optimize sample-level cache efficiency. These
systems treat the batch as a transient delivery artifact with no persistent
boundary, no atomic cross-rank visibility guarantee, and no mechanism for
deterministic replay. Dataset formats such as Petastorm~\cite{petastorm} and
MosaicML Streaming~\cite{mosaicmlstreaming} target static, pre-materialized
datasets and cannot linearize out-of-order arrivals from concurrent dynamic
producers into a durable step sequence. MegaScale-Data~\cite{megascaledata}
externalizes data construction but does not provide a persistent, checkpoint-aligned
batch history. Mixtera~\cite{boether2026mixtera} provides declarative mixture
control over static datasets but is a read-only layer with no transactional delivery
or fault tolerance semantics. General-purpose frameworks such as
Ray~\cite{moritz2018ray}, Spark~\cite{zaharia2012resilient}, and
Dask~\cite{rocklin2015dask} offer execution lineage, but lineage guarantees
recomputability rather than membership stability: recomputing a non-deterministic
batch produces a different result, which is insufficient for checkpoint recovery.
BatchWeave preserves the batch history as a stable, immutable record rather than
deriving it on demand.

\subsection{Lakehouse Architectures}
\label{sec:related:lakehouse}

Lakehouse systems such as Delta Lake~\cite{delta}, Apache Iceberg~\cite{iceberg},
and Lance~\cite{lance} use versioned metadata manifests over object storage to
coordinate concurrent writers via optimistic concurrency control. Magnus~\cite{magnus}
extends Iceberg for EB-scale ML data management with Git-like branching and
multimodal storage optimizations. BatchWeave is, to our knowledge, the first system
to repurpose this pattern as a streaming training data plane. BatchWeave adopts
the versioned-manifest transaction pattern of lakehouse systems, but goes beyond
conventional lakehouse semantics by making the Global Batch the fundamental unit
and enforcing training-specific consistency, including atomic all-rank
visibility, globally ordered batch progression, and checkpoint-aligned recovery
and reclamation. The DAC algorithm addresses a problem absent in offline
settings: sustaining high commit throughput as the manifest grows under
continuous concurrent production.

\section{Conclusion}
\label{sec:conclusion}

Modern LFM training demands more from the data pipeline than record delivery:
the Global Batch is the unit of distributed optimization, and its boundaries,
ordering, and lifecycle must align with checkpoint state. BatchWeave addresses
this by adapting versioned-manifest lakehouse design into a training-aware data
plane. It introduces the Transactional Global Batch, DAC for sustained
ingestion under manifest growth, durable producer state for exactly-once
recovery, and checkpoint-aligned reclamation tied to live checkpoints.
Implemented entirely as a client-side library with no server-side processes,
BatchWeave shows that object storage, combined with versioned manifests and
decentralized coordination, is a sufficient substrate for a training-aware data
plane.
 
\bibliographystyle{ACM-Reference-Format}
\bibliography{main}

\end{document}